\documentclass{sig-alternate-05-2015}

\usepackage{amsmath,amssymb,graphicx}
\usepackage{theorem}
\usepackage{here}
\usepackage{url}
\usepackage{comment}
\usepackage{algorithm}
\usepackage{algorithmicx}
\usepackage[noend]{algpseudocode}
\usepackage{graphicx}

\newcommand{\argmax}{\operatornamewithlimits{argmax}}

\begin{document}

%

\CopyrightYear{2016} 
\setcopyright{acmcopyright}
\conferenceinfo{KDD '16,}{August 13-17, 2016, San Francisco, CA, USA}
\isbn{978-1-4503-4232-2/16/08}\acmPrice{\$15.00}
\doi{http://dx.doi.org/10.1145/2939672.2939864}

\title{Scalable Partial Least Squares Regression on Grammar-Compressed Data Matrices}
%
%
%
%
%

\numberofauthors{4} 
%
\author{
%
%
\alignauthor
Yasuo Tabei \\
  \affaddr{Japan Science and Technology Agency, Japan} \\
  \email{tabei.y.aa@m.titech.ac.jp}
\alignauthor
Hiroto Saigo \\
  \affaddr{Kyushu University, Japan} \\
  \email{saigo@inf.kyushu-u.ac.jp}
\and
\alignauthor
Yoshihiro Yamanishi \\
  \affaddr{Kyushu University, Japan} \\
  \email{yamanishi@bioreg.kyushu-u.ac.jp}
\alignauthor
Simon J. Puglisi \\
  \affaddr{University of Helsinki, Finland} \\
  \email{simon.puglisi@cs.helsinki.fi}
}

\maketitle
\begin{abstract}

  With massive high-dimensional data now commonplace in research and industry,
  there is a strong and growing demand for more scalable computational techniques for data analysis 
  and knowledge discovery. Key to turning these data into knowledge is the ability to learn statistical
  models with high interpretability. Current methods for learning statistical models either produce
  models that are not interpretable or have prohibitive computational costs when applied to massive data.
  In this paper we address this need by presenting a scalable algorithm for partial least squares regression 
  (PLS), which we call compression-based PLS (cPLS), to learn predictive linear models with a high 
  interpretability from massive high-dimensional data.
  We propose a novel grammar-compressed representation of data matrices that supports fast row and column access 
  while the data matrix is in a compressed form. 
  The original data matrix is grammar-compressed and then the linear model in PLS is learned on the compressed 
  data matrix, which results in a significant reduction in working space, greatly improving scalability.  
  We experimentally test cPLS on its ability to learn linear models for classification, regression and feature 
  extraction with various massive high-dimensional data, and show that cPLS performs superiorly in terms 
  of prediction accuracy, computational efficiency, and interpretability.

\end{abstract}




\section{Introduction}

Massive data are now abundant throughout research and industry, in areas such as biology, chemistry, economics, digital libraries and data management systems.
In most of these fields, extracting meaningful knowledge from a vast amount of data is now the key challenge.
For example, to remain competitive, e-commerce companies need to constantly analyze huge data of user reviews and purchasing histories~\cite{McAuley13}.
In biology, 
detection of functional interactions of compounds and proteins is an important part in genomic drug discovery~\cite{Stockwell00,Dobson2004Chemical} and requires analysis of a huge number of chemical compounds~\cite{Chen09} 
 and proteins coded in fully sequenced genomes~\cite{UniProt10}. 
 There is thus a strong and growing demand for developing new, more powerful methods to make better use of massive data and to discover meaningful knowledge on a large scale.


Learning statistical models from data is an attractive approach for making use of massive high-dimensional data.
However, due to high runtime and memory costs, learning of statistical models from massive data --- especially models that have 
high interpretability --- remains a challenge. 


\begin{table*}[t]
\caption{Summary of scalable learning methods of linear models.}
\label{tab:related}
\label{methods}
{\small
  \begin{center}
    \setlength{\tabcolsep}{2pt}
  \begin{tabular}{c||c|c|c|c|c}
                                & Approach            & Compression Type & \# of parameters & Interpretability & Optimization \\ \hline
PCA-SL~\cite{Jolliffe86,Elgamal15} & Orthogonal rotation & Lossy & 2 & Limited & Stable \\
bMH-SL~\cite{Li11}                 & Hashing             & Lossy & 3 & Unable & Stable \\
SGD~\cite{Tsuruoka09, Duchi11} & Sampling     & -     & 1 & Limited & Unstable \\
\hline
\hline
cPLS (this study) & Grammar compression & Lossless & 1 & High & Stable \\
\end{tabular}
\end{center}
}
\end{table*}


{\em Partial least squares regression (PLS)} is a linear statistical model with latent features behind high-dimensional data~\cite{Rosipal06,Wold75,Wold01} that greedily finds
the latent features by optimizing the objective function under the orthogonal constraint. 
PLS is suitable for data mining, because extracted latent features in PLS provide a low-dimensional feature representation of the original data, making it easier for practitioners to interpret the results.
From a technical viewpoint, the optimization algorithm in PLS depends only on elementary matrix calculations of addition and multiplication. 
Thus, PLS is more attractive than other machine learning methods that are based on computationally burdensome mathematical programming and complex optimization solvers. 
In fact, PLS is the most common chemoinformatics method in pharmaceutical research.

However, applying PLS to massive high-dimensional data is problematic.
While the memory for the optimization algorithm in PLS depends only on the size of the corresponding data matrix, 
storing all high-dimensional feature vectors in the data matrix consumes a huge amount of memory, which 
limits large-scale applications of PLS in practice.
One can use {\em lossy compression} (e.g., PCA~\cite{Jolliffe86,Elgamal15} and $b$-bit minwise hashing~\cite{Gio99,Li10}) to compactly represent data matrices and then learn linear models on the compact data matrices~\cite{Li11}. 
However, although these lossy compression-based methods effectively reduce memory usage~\cite{Li11,Tabei13}, their drawback is that they cannot extract informative features from the learned models, because the original data matrices cannot be recovered from the compressed ones.


{\em Grammar compression}~\cite{Charikar05,Rytter03,Kieffer00} is a method of {\em lossless compression} (i.e., 
the original data can be completely recovered from grammar-compressed data) that
also has a wide variety of applications in string processing,
such as pattern matching~\cite{Yamamoto11}, edit-distance computation~\cite{Hermelin09}, and $q$-gram mining~\cite{Bille13}. 
Grammar compression builds a small context-free grammar that generates only the input data and 
is very effective at compressing sequences that contain many repeats.  
In addition, the set of grammar rules has a convenient representation as a forest 
of small binary trees, which enables us to implement various string operations without 
decompression. To date, grammar compression has been applied only to string (or sequence) data; however, 
as we will see, there remains high potential for application to other data representations.
A fingerprint (or bit vector) is a powerful representation of natural language texts~\cite{Manning99}, 
bio-molecules~\cite{Todeschini2002}, and images~\cite{Forsyth02}. 
Grammar compression is expected to be effective for compressing a set of fingerprints as well, 
because fingerprints belonging to the same class share many identical features.

\smallskip
{\em Contribution}. In this paper, we present a new scalable learning algorithm for PLS, which we call {\em lossless compression-based PLS (cPLS)}, to learn highly-interpretable predictive linear models 
from massive high-dimensional data. 
A key idea is to convert high-dimensional data with fingerprint representations into a set of sequences and then build grammar rules for representing the sequences in order to compactly store data matrices in memory. 
To achieve this, we propose a novel grammar-compressed representation of a data matrix capable of supporting row and column access {\em while the data matrix is in a compressed format}. 
The original data matrix is grammar-compressed, and then a linear model is learned on the compressed data matrix, which allows us to significantly reduce working space. 
cPLS has the following desirable properties: 
 \vspace{-0.5\baselineskip}           
{
\setlength{\leftmargini}{15pt}         
\begin{enumerate}
  \setlength{\itemsep}{0cm}      
  \setlength{\parskip}{0cm}      
  \setlength{\itemindent}{0pt}   
  \setlength{\labelsep}{3pt}     
\item {\bf Scalability}: cPLS is applicable to massive high-dimensional data.
\item {\bf Prediction Accuracy}: cPLS can achieve high prediction accuracies for both classification and regression.
\item{\bf Usability}: cPLS has only one hyper parameter, which enhances the usability of cPLS.
\item {\bf Interpretability}: Unlike lossy compression-based methods, cPLS can extract features reflecting the correlation structure between data and class labels/response variables.
\end{enumerate}
}
 \vspace{-0.5\baselineskip}           
We experimentally test cPLS on its ability to learn linear models for classification, regression and feature extraction with 
various massive high-dimensional data, and show that cPLS performs superiorly in terms of prediction accuracy, computational efficiency, and interpretability.

\section{Literature Review}\label{sec:related}
Several efficient algorithms have been proposed for learning linear models on a large scale.
We now briefly review the state of the art, which is also summarized in Table~\ref{tab:related}.

{\em Principal component analysis (PCA)}~\cite{Jolliffe86} is a widely used machine learning tool, and 
is a method of lossy compression, i.e., the original data cannot be recovered from compressed data. 
There have been many attempts to extend PCA~\cite{Tipping00,Scholkopf98} and present a scalable PCA in distributed settings for analyzing big data~\cite{Elgamal15}. 
For classification and regression tasks, a data matrix is compressed by PCA, and linear models are learned on the compressed data matrix by a {\em supervised learning method (SL)}, which is referred to as {\em PCA-SL}.
Despite these attempts, PCA and its variants do not look at the correlation structure between data and output variables (i.e., class labels/response variables), which 
results in not only the inability of feature extractions in PCA but also the inaccurate predictions by PCA-SL. 

Li et al.~\cite{Li11} proposed a compact representation of fingerprints for learning linear models by applying {\em $b$-bit minwise hashing (bMH)}.
A $d$-dimensional fingerprint is conceptually equivalent to the set $s_{i} \subset \{1,...,d\}$ that contains element $i$ if and only if the $i$-th bit in the fingerprint is $1$. Li et al.'s method works as follows.
We first pick $h$ random permutations $\pi_i$, $i=1,..,h$, each of which maps $[1,d]$ to $[1,d]$.
We then apply a random permutation $\pi$ on a set $s_i$, compute the minimum element as $\min(\pi(s_i))$, 
and take as a hash value its lowest $b$ bits.
Repeating this process $h$ times generates $h$ hash values of $b$ bits each.
Expanding these $h$ values into a ($2^b \times h$)-dimensional fingerprint with exactly $h$ $1$'s builds a compact representation of the original fingerprint.

Linear models are learned on the compact fingerprints by SL, which is referred to as {\em bMH-SL}.
Although bMH-SL is applicable to large-scale learning of linear models,
bMH is a method of lossy compression and cannot extract features from linear models learned by SL.
Other hashing-based approaches have been proposed such as Count-Min sketch~\cite{Cormode05}, Vowpal Wabbit~\cite{Weinberger09}, and Hash-SVM~\cite{Mu14}.
However, like bMH-SL, these algorithms cannot extract features, which is a serious problem in practical applications.

{\em Stochastic gradient descent (SGD)}~\cite{Duchi11, Tsuruoka09} is a computationally efficient algorithm for learning linear models on a large-scale.
SGD samples $\nu$ feature vectors from an input dataset and computes the gradient vector from the sampled feature vectors.
The weight vector in linear models is updated using the gradient vector and the learning rate $\mu$, and
this process is repeated until convergence.
Unfortunately however, learning linear models using SGD is numerically unstable, resulting in low prediction accuracy.
This is because SGD has three parameters ($\nu$, $\mu$, and $C$) that must be optimized if high classification accuracy 
is to be attained.
Online learning is a specific version of SGD that loads an input dataset from the beginning and updates the weight vector in a linear model for each feature vector. 
{\em AdaGrad}~\cite{Duchi11} is an efficient online learning that automatically tunes parameters of $\nu$ and $\mu$ in SGD. 
Although online learning is space-efficient (owing to its online nature), it is also numerically unstable. 
Even worse, AdaGrad is applicable only to differentiable loss functions, 
which limits its applicability to simple linear models, e.g., SVM and logistic regression, 
making the learned model difficult to interpret.

Despite the importance of scalable learning of interpretable linear models, 
no previous work has been able to achieve high prediction accuracy for classification/regression tasks and high interpretability of the learned models.
We present a scalable learning algorithm that meets both these demands and
is made possible by learning linear models on grammar-compressed data in the framework of PLS. Details of the proposed method are presented in the next section.

\section{Grammar Compression}
\begin{figure}
\begin{center}
\begin{tabular}{c}
\includegraphics[width=0.4\textwidth]{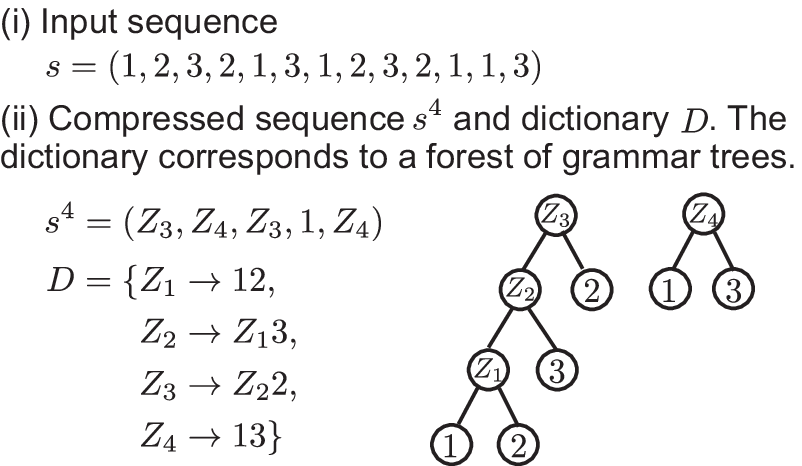}
\end{tabular}
\end{center}
\vspace{-1cm}
\caption{Illustration of grammar compression.}
\label{fig:grammar}
\end{figure}

Given a sequence of integers $S$, a {\em grammar-compressor} generates a context-free grammar (CFG)
that generates $S$ and only $S$.
The grammar consists of a set of rules\footnote{In this paper we assume without loss of generality that the grammar is in Chomsky Normal Form.}.
Each rule is of the form $Z_i\to ab$. 
Symbols that appear on the left-hand side of any rule are called {\em non-terminals}.
The remaining symbols are called {\em terminals}, all of which are present in the input sequence.
Informally, a rule $Z_i\to ab$ indicates that on the way to recovering the original 
sequence from its grammar-compressed representation, occurrences of the symbol $Z_i$ should be replaced 
by the symbol pair $ab$ (the resulting sequence may then be subject to yet more replacements).
A data structure storing a set of grammar rules is called a {\em dictionary} and 
is denoted by $D$. Given a non-terminal, the dictionary supports access to the symbol 
pair on the right-hand of the corresponding grammar rule, i.e., $D[Z_i]$ returns $ab$ 
for rule $Z_i\to ab$. 
The original sequence can be recovered from the compressed sequence and $D$. 
The set of grammar rules in $D$ can be represented as a forest of (possibly small)
binary trees called {\em grammar trees}, where each node and its left/right children 
correspond to a grammar rule. 
See Figure~\ref{fig:grammar} for an illustration.

The size of a grammar is measured as the number of rules plus the size of compressed sequence.
The problem of finding the minimal grammar producing a given string is known to be NP-complete~\cite{Charikar05}, 
but several approximation algorithms exist that produce grammars that are small in practice
(see, e.g.,~\cite{Rytter03,Larsson99,Kieffer00}).
Among these is the simple and elegant Re-Pair~\cite{Larsson99} algorithm, 
which we review next.




\subsection{Re-Pair Algorithm}
The Re-Pair grammar compression algorithm by Larsson and Moffat~\cite{Larsson99}
builds a grammar by repeatedly replacing the most frequent symbol pair in an 
integer sequence with a new non-terminal. Each iteration of the algorithm consists
of the following two steps:
(i) find the most frequent pair of symbols in the current sequence, and then 
(ii) replace the most frequent pair with a new non-terminal symbol, generating a new grammar rule and a new 
(and possibly much shorter) sequence. Steps (i) and (ii) are then applied to the new sequence and iterated 
until no pair of adjacent symbols appears twice. 

Apart from the dictionary $D$ that stores the rules as they are generated, Re-Pair maintains a hash table
and a priority queue that together allow the most frequent pair to be found in each iteration. The hash table, 
denoted by $H$, holds the frequency of each pair of adjacent symbols $ab$ in the current sequence, i.e., $H:ab\to \textbf{N}$. 
The priority queue stores the symbol pairs keyed on frequency and allows the most frequent symbol to be found in step~(i).
In step~(ii), a new grammar rule $Z_1\to ab$ is generated where $ab$ is the most frequent symbol 
pair and $Z_1$ is a new non-terminal not appearing in a sequence. The rule is stored in the dictionary $D$. 
Every occurrence of $ab$ in the sequence is then replaced by $Z_1$, generating a new, shorter 
sequence. This replacement will cause the frequency of some symbol pairs to change,
so the hash table and priority queue are then suitably updated.


Let $s^c$ denote a sequence generated at $c$-th iteration in the Re-Pair algorithm. 
For input sequence $s$ in Figure~\ref{fig:grammar}, the most frequent pair of symbols is $12$. 
Thus, we generate rule $Z_1\to 12$ to be added to the dictionary $D$ and replace all the occurrences of $12$ by non-terminal $Z_1$ in $s$. 
After four iterations, the current sequence $s^4$ has no repeated pairs, 
and thus the algorithm stops. 
Dictionary $D$ has four grammar rules that correspond to a forest of two small trees. 

As described by Larsson and Moffat~\cite{Larsson99}, Re-Pair can be implemented to run in linear time in the length 
of the input sequence, but it requires the use of several heavyweight data structures to 
track and replace symbol pairs. The overhead of these data structures (at least 128 bits per 
position) prevents the algorithm from being applied to long sequences, such as the large data matrices.


Another problem that arises when applying Re-Pair to long sequences is the memory required for storing 
the hash table: a considerable number of symbol pairs appear twice in a long sequence, and
the hash table stores something for each of them, consuming large amounts of memory. 

In the next section, we present scalable Re-Pair algorithms that achieve both 
space-efficiency and fast compression time on large data matrices. 
Specifically, our algorithms need only constant working space. 

\section{Our Grammar-Compressed Data Matrix}

Our goal is to obtain a compressed representation of a data matrix ${\textbf X}$ of $n$ rows and $d$ columns.
Let $x_i$ denote the $i$th row of the matrix represented as a fingerprint (i.e. binary vector).
An alternative view of a row that will be useful to us is as a sequence of integers 
$s_i=(p_1,p_2,...,p_m)$, $p_1<p_2<\cdots < p_m$, where $p_i \in s_i$ if and only if $x_i[p_i] = 1$.
In other words the sequence $s_i$ indicates the positions of the 1 bits in $x_i$.

In what follows we will deal with a differentially encoded form of $s_i$ in which the difference 
for every pair of adjacent elements in $s_i$ is stored, i.e., $s_i=(p_1,p_2,...,p_m)$ is encoded 
as $s_{gi}=(p_1,p_2-p_1,p_3-p_2,...,p_m-p_{m-1})$. This differential encoding tends to increase 
the number of repeated symbol pairs, which allows the sequences $s_{gi}$ to be more effectively 
compressed by the Re-Pair algorithm. 
A grammar compressor captures the underlying correlation structure of data matrices:
by building the same grammar rules for the same (sequences of) integers, it effectively compresses data matrices with many repeated integers. 

%

\subsection{Re-Pair Algorithms in Constant Space}
We now present two ideas to make Re-Pair scalable without seriously deteriorating its compression performance. 
Our first idea is to modify the Re-Pair algorithm to identify top-$k$ frequent symbol pairs in all rows $s^c_{gi}$ in 
step~(i) and replace all the occurrences of the top-$k$ symbol pairs in all rows $s^c_{gi}$ in step~(ii), 
generating new $k$ grammar rules and new rows $s^{c+1}_{gi}$.
This new replacement process improves scalability by reducing the number of iterations required by roughly a factor of $k$. 

Since we cannot replace both frequent symbol pairs $ab$ and $bc$ in triples $abc$ in step~(ii), we replace the 
first appearing symbol pair $ab$, preferentially. However, such preferential replacement can generate a replacement 
of a pair only once and can add redundant rules to a dictionary, adversely affecting compression performance. 
To overcome this problem, we replace the first and second appearances of each frequent pair at the same time and 
replace the next successive appearance of the frequent pair as usual, which guarantees generating grammar rules that appear
at least twice. 

Our second idea is to reduce the memory of the hash table by removing infrequent symbol pairs. 
Since our modified Re-Pair algorithm can work storing compressed sequences $s_{gi}^c$ at each iteration $c$ in a secondary storage device, 
the hash table consumes most of the memory in execution. 
Our modified Re-Pair generates grammar rules from only top-$k$ frequent symbol pairs in the hash table, 
which means only frequent symbol pairs are expected to contribute to the compression. 
Thus, we remove infrequent symbol pairs from the hash table by leveraging 
the idea behind stream mining techniques originally proposed in~\cite{Karp03,DemaineLM02,MankuM12} for finding frequent items 
in data stream. 
Our method is a counter-based algorithm that computes the frequency of each symbol pair and removes infrequent ones 
from the hash table at each interval in step~(i). We present two Re-Pair algorithms using lossy counting and frequency counting 
for removing infrequent symbol pairs from the hash table. 
We shall refer to the Re-Pair algorithms using lossy counting and frequency counting as Lossy-Re-Pair and Freq-Re-Pair, respectively. 

\subsection{Lossy-Re-Pair}
The basic idea of lossy counting is to divide a sequence of symbols into intervals of fixed length and 
keep symbol pairs in successive intervals in accordance with their appearance frequencies in a hash table. 
Thus, if a symbol pair has appeared $h$ times in the previous intervals, it is going to be kept in the 
next $h$ successive intervals. 

Let us suppose a sequence of integers made by concatenating all rows $s_{gi}$ of ${\bf X}$ and let $N$ be the length of the sequence. 
We divide the sequence into intervals of fixed-length $\ell$. Thus, the number of intervals is $N/\ell$. 
We use hash table $H$ for counting the appearance frequency of each symbol pair in the sequence. 
If symbol pair $ab$ has count $H(ab)$, it is ensured that $ab$ is kept in hash table $H$ until the next $H(ab)$-th interval. 
If symbol pair $ab$ first appears in the $q$-th interval, $H(ab)$ is initialized as $qN/\ell + 1$, which ensures that $ab$ is kept until at least the next interval,
i.e., the $(qN/\ell+1)$-th interval. 
Algorithm~\ref{alg:lossy} shows the pseudo-code of lossy counting. 

The estimated number of symbol pairs in the hash table is $O(\ell)$~\cite{MankuM12}, 
resulting in $O(\ell \log\ell)$ bits consumed by the hash table. 

\begin{algorithm}
\caption{Lossy counting. $H$: hash table, $N$: length of an input string at a time point, 
$\ell$: length of each interval. Note that lossy counting can be used in step~(i) in the Re-Pair algorithm.}
\label{alg:lossy}
\begin{algorithmic}[1]
\State Initialize $N=0$ and $\Delta=0$
\Function{LossyCounting}{$a,b$}
\State $N=N+1$
\If{$H(ab) \neq 0$}
\State $H(ab)=H(ab)+1$
\Else
\State $H(ab)=\Delta + 1$
\EndIf
\If{$\lfloor\frac{N}{\ell} \rfloor \neq \Delta$}
\State $\Delta=\lfloor \frac{N}{\ell} \rfloor$
\For{each symbol pair $ab$ in $H$}
\If{$H(ab)<\Delta$}
\State Remove $ab$ from $H$
\EndIf
\EndFor
\EndIf
\EndFunction
\end{algorithmic}
\end{algorithm}

\subsection{Freq-Re-Pair}
The basic idea of frequency counting is to place a limit, $v$, on the maximum number of symbol pairs 
in hash table $H$ and then keep only the most frequent $v$ symbol pairs in $H$.
Such frequently appearing symbol pairs are candidates to be replaced by new non-terminals, which generates 
a small number of rules. 

The hash table counts the appearance frequency for each symbol pair in step~(i) of the Re-Pair algorithm. 
When the number of symbol pairs in the hash table reaches $v$, Freq-Re-Pair removes the bottom 
$\epsilon$ percent of symbol pairs with respect to frequency. We call $\epsilon$ the vacancy rate. 
Algorithm~\ref{alg:freq} shows the pseudo-code of frequency counting. 
The space consumption of the hash table is $O(v \log{v})$ bits. 

\begin{algorithm}
\caption{Frequency counting. $H$: hash table, $|H|$: number of symbol pairs in $H$, 
$v$: the maximum number of symbol pairs in $H$, 
$\epsilon$: vacancy rate. 
Note that frequency counting can be used in step~(i) in the Re-Pair algorithm.}
\label{alg:freq}
\begin{algorithmic}[1]
\Function{FrequencyCounting}{$a,b$}
\If{$H(ab) \neq 0$}
\State $H(ab)=H(ab)+1$
\Else
\If{$|H| \geq v$}
\While{$v(1-\epsilon/100)<|H|$}
\For{each symbol pair $a^\prime b^\prime$ in H}
\State $H(a^\prime b^\prime)=H(a^\prime b^\prime)-1$
\If{$H(a^\prime b^\prime)=0$}
\State Remove $a^\prime b^\prime$ from $H$
\EndIf
\EndFor
\EndWhile
\EndIf
\State $H(ab)=1$
\EndIf
\EndFunction
\end{algorithmic}
\end{algorithm}

\section{Direct Access to Row and Column} \label{sec:rowcol}
In this section, we present algorithms for directly accessing rows and columns of a grammar-compressed data matrix, 
which is essential for us to be able to apply PLS on the compressed matrix in order to learn linear regression models.

\subsection{Access to Row}
Accessing the $i$-th row corresponds to recovering the original $s_i$ from grammar-compressed $s_{gi}^c$.
We compute this operation by traversing the grammar trees. 
For recovering the $i$-th row $s_{i}$, 
we start traversing the grammar tree having a node of the $q$-th symbol $s_{gi}^c[q]$ as a root for each $q$ from $1$ to $|s_{gi}^c|$.
Leaves encountered in the traversal must have integers in sequence $s_{gi}$, which 
allows us to recover $s_{gi}$ via tree traversals, starting from the nodes with non-terminal $s^c_{gi}[q]$ for each $q\in [1,|s_{gi}^c|]$.
We recover the original $i$-th row $s_i$ from $s_{gi}$ by cumulatively adding integers in $s_{gi}$ from $1$ to $|s_{gi}|$, 
i.e, $s_i[1]=s_{gi}[1]$, $s_i[2]=s_{gi}[2]+s_i[1]$,...,$s_i[|s_{gi}|]=s_{gi}[|s_{gi}|]+s_i[|s_{gi}|-1]$. 

\subsection{Access to Column}
Accessing the $j$-th column of a grammar-compressed data matrix requires us to obtain a set of row identifiers $R$ such that $x_{ij}=1$ for 
$i\in [1,n]$, i.e., $R=\{i\in [1,n]; x_{ij}=1 \}$. 
This operation enables us to compute the transpose ${\textbf X}^\intercal$ from ${\textbf X}$ in compressed format, which is used in the optimization algorithm of PLS. 

$P[Z_i]$ stores a summation of terminal symbols as integers at the leaves under the node corresponding to terminal symbol $Z_i$ in a grammar tree.
For example, in Figure~\ref{fig:grammar}, $P[Z_1]=3$, $P[Z_2]=6$, $P[Z_3]=8$ and $P[Z_4]=4$. 
$P$ can be implemented as an array that is randomly accessed from a given non-terminal symbol.
We shall refer to $P$ as the weight array. 
The size of $P$ depends only on the grammar size. 

The $j$-th column is accessed to check whether or not $x_{ij} = 1$ in compressed sequence $s^c_{gi}$, for each $i \in [1,n]$. 
We efficiently solve this problem on grammar-compressed data matrix by using the weight array $P$. 
Let $u_q$ store the summation of weights from the first symbol $s^c_{gi}[1]$ to the $q$-th symbol $s^c_{gi}[q]$, i.e., $u_q=P[s^c_{gi}[1]]+P[s^c_{gi}[2]]+\cdots+P[s^c_{gi}[q]]$, 
and let $u_0=0$.
If $u_{q}$ is not less than $j$, the grammar tree with the node corresponding to a symbol $s^c_{gi}[q]$ as a root can encode $j$ at a leaf.
Thus, we traverse the tree in depth-first order from the node corresponding to symbol $s^c_{gi}[q]$ as follows.
Suppose $Z=s^c_{gi}[q]$ and $u=u_{q-1}$. 
Let $Z_\ell$ (resptively $Z_r$) be $a$ (respectively $b$) of $Z\to ab$ in $D$. 
(i) if $j < u$, we go down to the left child in the tree; 
(ii) otherwise, i.e., $j \geq u$, we add $P[Z_\ell]$ to $u$ and go down to the right child.
We continue the traversal until we reach a leaf. 
If $s=j$ at a leaf, this should be $x_{ij}=1$ at row $i$; thus we add $i$ to solution set $R$. 
Algorithm~\ref{alg:column} shows the pseudo-code for column access. 

\begin{algorithm}[t]
\caption{Access to the $j$-th column on grammar-compressed data matrix. $R$: solution set of row identifiers $i$ at column $j$ s.t. $x_{ij}=1$.}
\label{alg:column}
\begin{algorithmic}[1]
\Function{AccessColumn}{$j$}
\For{$i$ in $1..n$}
\State $u_{0}=0$
\For{$q$ in $1..|s^c_{gi}|$}
\State $u_{q}=u_{q-1}+P[S^c_{gi}[q]]$
\If{$j \leq u_{q}$}
\State {\sc Recursion}$(i,j,s^c_{gi}[q],u_{q-1})$
\State break
\EndIf
\EndFor
\EndFor
\EndFunction
\end{algorithmic}
\begin{algorithmic}[1]
\Function{Recursion}{$i$,$j$,$Z$,$u$}
\If{$Z$ is a terminal symbol}
\If{$u+Z=j$}
\State Add $i$ to $R$
\EndIf
\State return
\EndIf
\State Set $Z_l$ (resp. $Z_r$) as $a$ (resp. $b$) of $Z\to ab$ in $D$
\If{$u+P[Z_{l}] > j$}
\State {\sc Recursion}($i$,$j$,$Z_l$,$u$) \Comment{Go to left child}
\Else
\State {\sc Recursion}($i$,$j$,$Z_r$,$u+P[Z_l]$) \Comment{Go to right child}
\EndIf
\EndFunction
\end{algorithmic}
\end{algorithm}

\section{cPLS}
In this section we present our cPLS algorithm for learning PLS on grammar-compressed data matrices. 
We first review the PLS algorithm on uncompressed data matrices. 
NIPALS~\cite{Wold75} is the conventional algorithm for learning PLS and requires the deflation of the data matrix involved.
We thus present a non-deflation PLS algorithm for learning PLS on compressed data matrices. 

\subsection{NIPALS}
Let us assume a collection of $n$ data samples and their output variables $(x_1,y_1),$ $(x_2,y_2),...,(x_n,y_n)$ where $y_i \in \Re$. 
The output variables are assumed to be centralized as $\sum_{i=1}^{n}y_i=0$. Denote by $y\in \Re^n$ the vector of all the 
training output variables, i.e., $y=(y_1,y_2,...,y_n)^\intercal$. 

The regression function of PLS is represented by the following special form, 
\[
  f(x) = \sum_{i=1}^m \alpha_i w_i^\intercal x, 
\]
where the $w_i$ are weight vectors reducing the dimensionality of $x$; they satisfy the following orthogonality condition:
\begin{eqnarray} \label{eq:ortho}
  w_i^\intercal\textbf{X}^\intercal\textbf{X}w_j = \left\{
               \begin{array}{cl}
                 1 & i = j \\
                 0 & i \neq j
               \end{array}.
\right.
\end{eqnarray}

We have two kinds of variables $w_i$ and $\alpha_i$ to be optimized. 
Denote by ${\textbf W} \in \Re^{d\times m}$ the weight matrix $i$-th column of which is weight 
vector $w_i$, i.e., ${\textbf W}=(w_1,w_2,...,w_m)$. 
Let $\alpha \in \Re^m$ be a vector whose $i$-th element is $\alpha_i$, i.e., $\alpha=(\alpha_1,\alpha_2,...,\alpha_m)^\intercal$.
Typically, ${\textbf W}$ is first optimized and then $\alpha$ is determined 
by minimizing the least squares error without regularization, 
\begin{eqnarray} 
\label{eq:alpha}
  \min_{\alpha} ||y-\textbf{X}\textbf{W}\alpha||^2_2.
\end{eqnarray}
By computing the derivative of equation (\ref{eq:alpha}) with respect to $\alpha$ and 
setting it to zero, $\alpha$ is obtained as follows:
\begin{eqnarray}\label{eq:sol}
  \alpha = (\textbf{W}^\intercal\textbf{X}^\intercal\textbf{X}\textbf{W})^{-1}\textbf{W}^\intercal \textbf{X}^\intercal y.
\end{eqnarray}

The weight vectors are determined by the following greedy algorithm. 
The first vector $w_1$ is obtained by maximizing the squared covariance between the mapped feature $\textbf{X}w$ and 
the output variable $y$ as follows: $w_1 = \argmax_{w} cov^2(\textbf{X}w,y)$,
subject to $w^\intercal \textbf{X}^\intercal \textbf{X} w=1$, 
where $cov(\textbf{X}w,\textbf{y})=y^\intercal \textbf{X}w$. 
The problem can be analytically solved as 
$w_1=\mathbf{X}^\intercal y$.

For the $i$-th weight vector, the same optimization problem is solved with additional constraints to maintain orthogonality, 
\begin{eqnarray} 
\label{eq:cov-i}
  w_i = \argmax_{w} cov^2(\textbf{X}w,y),
\end{eqnarray}
subject to $w^\intercal \textbf{X}^\intercal \textbf{X} w=1$, $w^\intercal \textbf{X}^\intercal\textbf{X}^\intercal w_j=0$, $j=1,...,i-1$. 
The optimal solution of this problem cannot be obtained analytically, but NIPALS solves it indirectly. 
Let us define the $i$-th latent vector as $t_i=\textbf{X}w_i$.
The optimal latent vectors $t_i$ are obtained first and the corresponding $w_i$ is obtained later. 
NIPALS performs the deflation of design matrix $\textbf X$ to ensure the orthogonality between latent components $t_i$ as follows, $\textbf{X} = \textbf{X} - t_it_i^\intercal \textbf{X}$.
Then, the optimal solution has the form, $w_i = \mathbf{X}^\intercal y$.

Due to the deflation, ${\mathbf X} = {\mathbf X} - t_it_i^\intercal {\mathbf X}$, 
NIPALS completely destroys the structure of ${\mathbf X}$.
Thus, it cannot be used for learning PLS on grammar-compressed data matrices. 



\subsection{cPLS Algorithm}\label{sec:cpls}
 \begin{table*}[t]
\caption{Summary of datasets.}
\label{tab:statistics}
\begin{center}
{\footnotesize
\begin{tabular}{l|rrrrr}
Dataset & Label type & Number & Dimension & \#nonzeros & Memory (mega bytes) \\
\hline 
Book-review & binary & 12,886,488 & 9,253,464 & 698,794,696 & 2,665 \\
Compound & binary & 42,682 & 52,099,292 & 914,667,811 & 3,489 \\
Webspam & binary & 350,000 & 16,609,143 & 1,304,697,446 & 4,977 \\
CP-interaction & binary  & 216,121,626 & 3,621,623 & 32,831,736,508 & 125,243 \\
CP-intensity & real & 1,329,100 & 682,475 & 28,865,055,991 &  110,111 \\
\end{tabular}
}
\end{center}
\end{table*}

\begin{figure*}
\begin{center}
\begin{tabular}{ccccc}
{\bf Book-review} & {\bf Compound} & {\bf Webspam} & {\bf CP-interaction} & {\bf CP-intensity} \\
\includegraphics[width=0.18\textwidth]{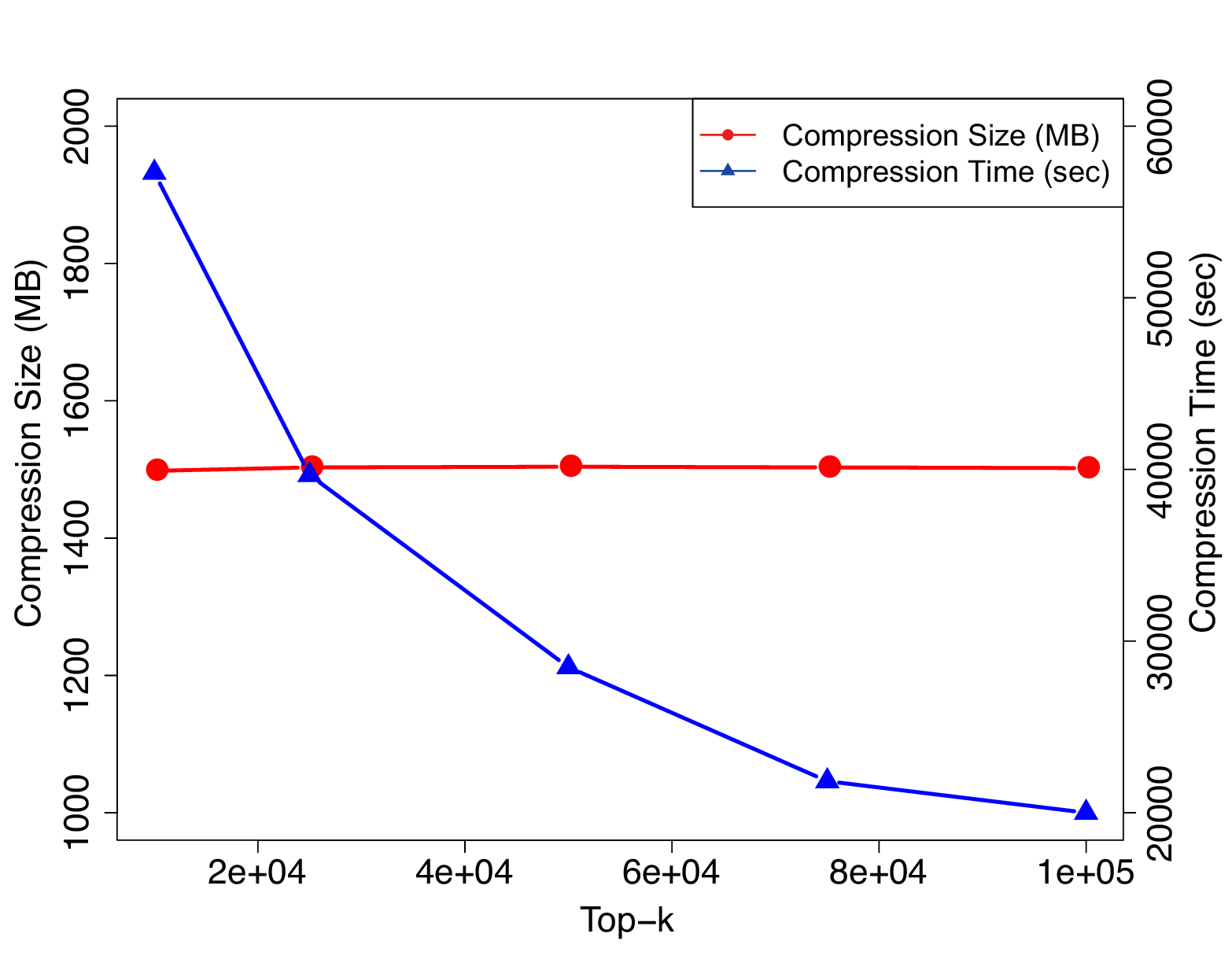} &
\includegraphics[width=0.18\textwidth]{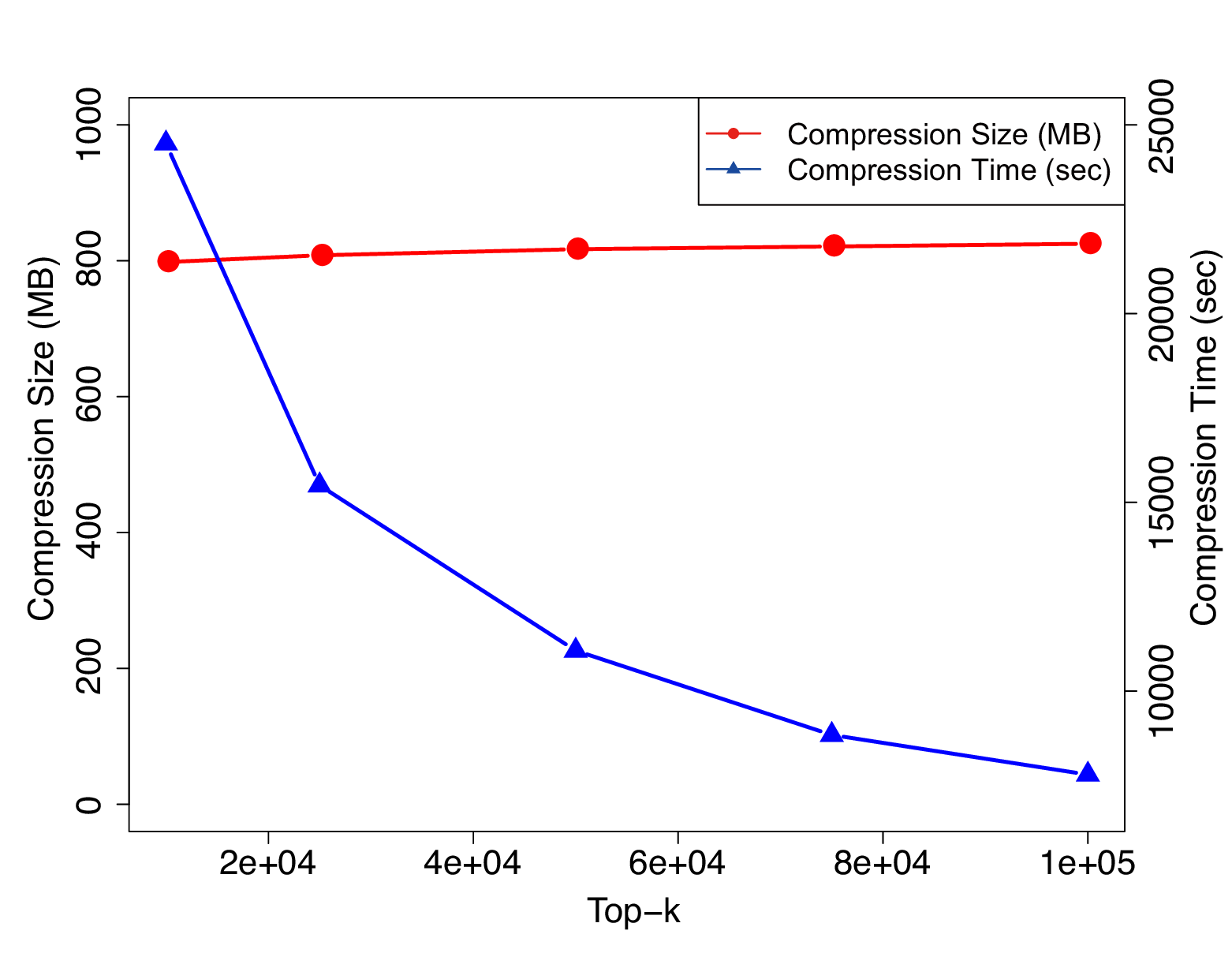} &
\includegraphics[width=0.18\textwidth]{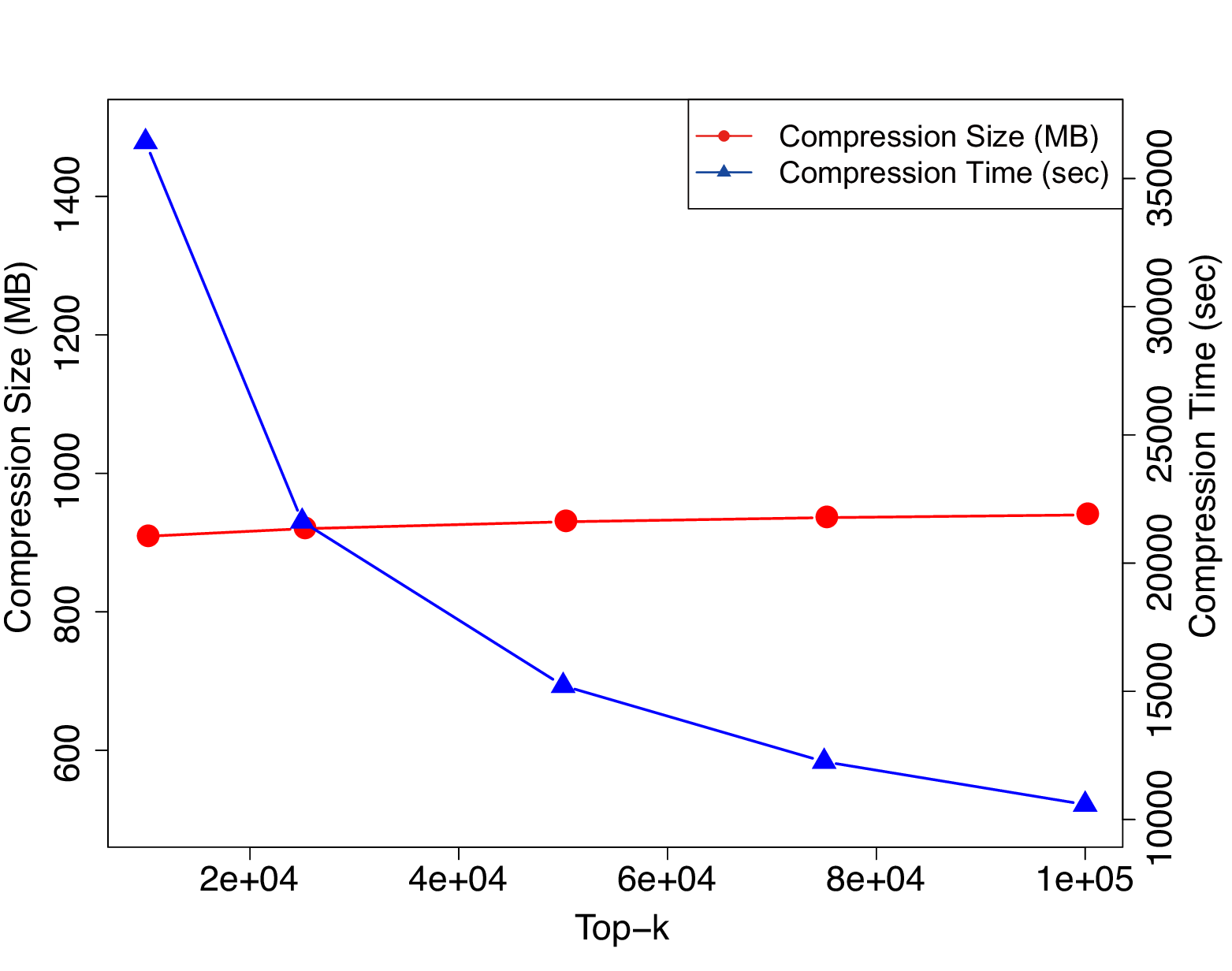} &
\includegraphics[width=0.18\textwidth]{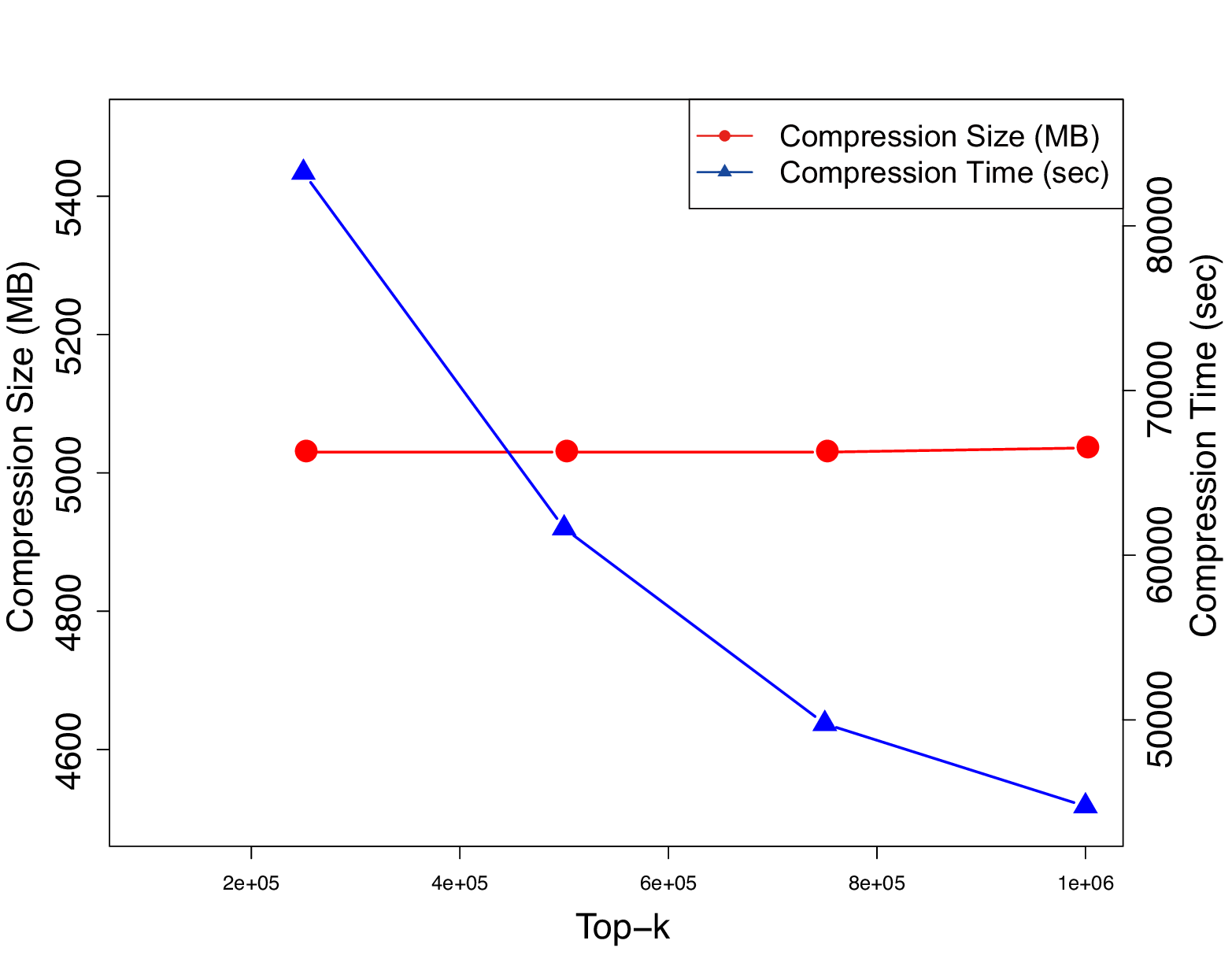} &
\includegraphics[width=0.18\textwidth]{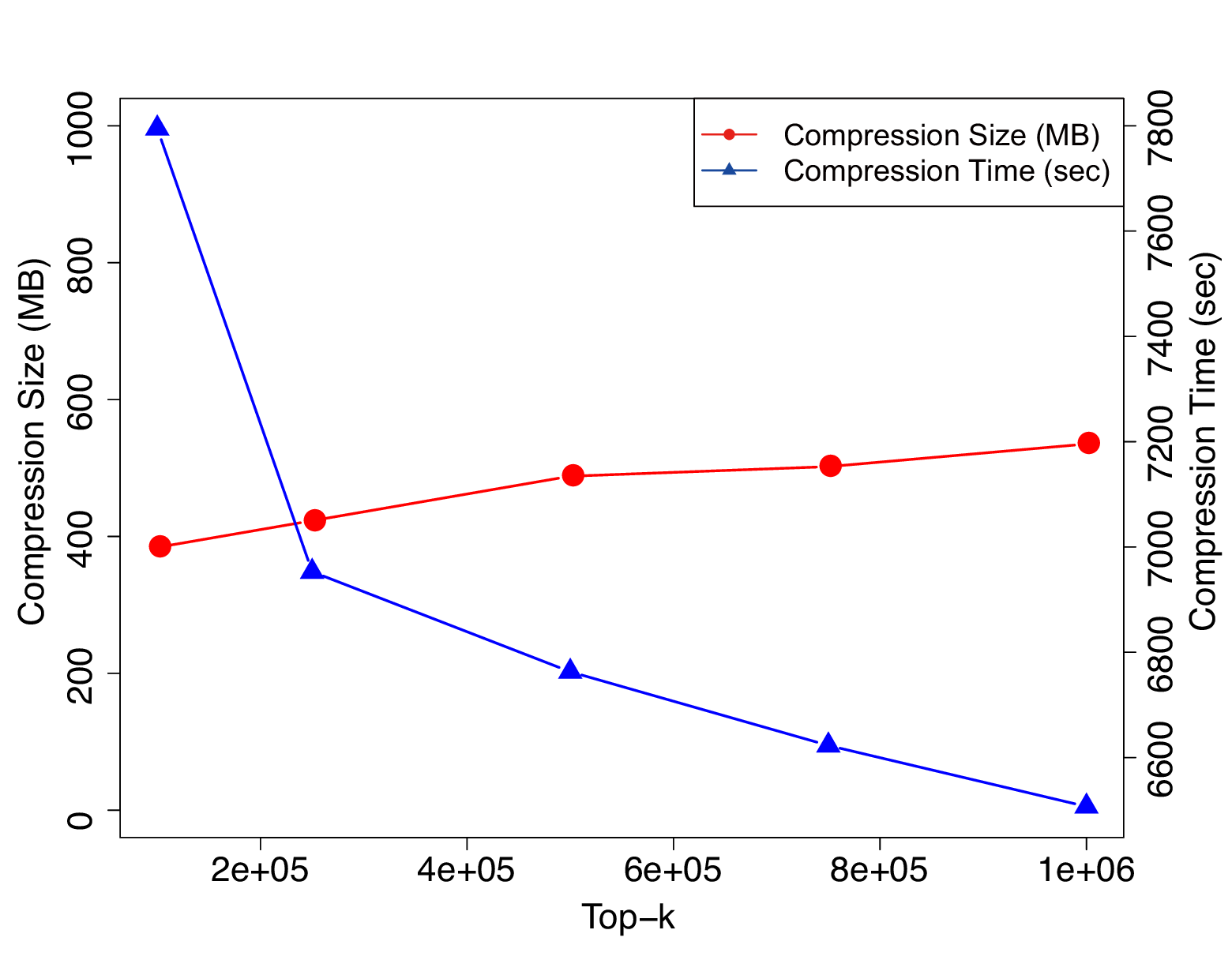} 
\end{tabular}
\end{center}
\vspace{-0.9cm}
\caption{Compression size in mega bytes (MB) and compression time in seconds (sec) for various top-$k$}
\label{fig:topk}
\end{figure*}

We present a non-deflation PLS algorithm for learning PLS on grammar-compressed data matrices. 
Our main idea here is to avoid deflation by leveraging the connection between NIPALS~\cite{Wold75} and 
the Lanczos method~\cite{Lanczos1950} which was 
originally proposed for recursive fitting of residuals without changing the structure of a data matrix.

We define residual $r_{i+1}=(r_i - (y^\intercal t_{i-1})t_{i-1})$ that is initialized as $r_1=y$. 
The $i$-th weight vector is updated as
$w_i = {\textbf X}^\intercal (r_{i-1} - (y^\intercal t_{i-1})t_{i-1})$, 
which means $w_i$ can be computed without deflating the original data matrix $\textbf{X}$. 
The $i$-th latent vector is computed as $t_i={\textbf X}w_i$ and is orthogonalized by applying 
the Gram-Schmidt orthogonalization to the $i$-th latent vector $t_i$ and 
previous latent vectors $t_1$,$t_2$,...,$t_{i-1}$ as follows, 
$t_i = (\textbf{I} - \textbf{T}_{i-1}\textbf{T}_{i-1}^\intercal)\textbf{X}w_i$, 
where $\textbf{T}_{i-1}=(t_1,t_2,...,t_{i-1}) \in \Re^{n\times (i-1)}$. 
The non-deflation PLS algorithm updates the residual $r_i$ instead of deflating ${\textbf X}$, thus
enabling us to learn PLS on grammar-compressed data matrices. 

cPLS is the non-deflation PLS algorithm that learns PLS on grammar-compressed data matrices. 
The input data matrix is grammar-compressed and then the PLS is learned on the compressed data matrix by the non-deflation PLS algorithm. 
Our grammar-compressed data matrix supports row and column accesses directly on the compressed format for 
computing matrix calculations of addition and multiplication, which enables us to learn PLS by using the non-deflation PLS algorithm. 
Let ${\textbf X}_G$ be the grammar-compressed data matrix of ${\textbf X}$. 
Algorithm~\ref{alg:1} shows the pseudo-code of cPLS. 
Since our grammar-compression is lossless, the cPLS algorithm on grammar-compressed data matrices learns the same model as 
the non-deflation PLS algorithm on uncompressed data matrices and so achieves the same prediction accuracy.  

\begin{algorithm}[h]
\caption{The cPLS algorithm. ${\textbf X}_G$: the grammar-compressed data matrix of ${\textbf X}$.}
\label{alg:1} 
\begin{algorithmic}[1]
\State $r_1=y$ 
\For{$i=1,...,m$}
\State $w_i = \textbf{X}_G^\intercal r_i$ \label{line:fec} \Comment{access to column} 
\If{$i=1$}
\State $t_1=\textbf{X}_G w_i$ \Comment{access to row}
\Else
\State $t_i=(\textbf{I}-\textbf{T}_{i-1}\textbf{T}_{i-1}^\intercal)\textbf{X}_Gw_{i}$ \Comment{access to row}
\EndIf
\State $t_i=t_i/||t_i||_2$
\State $r_{i+1}=r_i-(y^\intercal t_i)t_i$
\EndFor
\State Compute the coefficients $\alpha$ using equation~(\ref{eq:sol}).
\end{algorithmic}
\end{algorithm}

We perform feature extraction after line~\ref{line:fec} at each iteration in Algorithm~\ref{alg:1}. 
The features corresponding to the top-$u$ largest weights $w_i$ are extracted. 
Due to the orthogonality condition~(\ref{eq:ortho}), the extracted features give users a novel insight for analyzing data, 
which is shown in Section~\ref{sec:exp}.

The cPLS algorithm has three kinds of variables to be optimized: $w_i$, $r_i$, and $t_i$. 
The memory for $w_m$ is $O(md)$ and the memory for $t_m$ and $r_i$ is $O(mn)$.
Thus, the total memory for the variables in cPLS is $O(m\min(n,d))$ highly depending on parameter $m$.
The parameter $m$ controls the amount of fitting of the model to the training data and is typically chosen to optimize the cross validation error. 
Since the cPLS algorithm learns the model parameters efficiently, $m$ can be set to a small value, which results in overall space-efficiency.

\section{Experiments}\label{sec:exp}

\begin{figure*}[t]
\begin{center}
\begin{tabular}{c}
\includegraphics[width=0.72\textwidth]{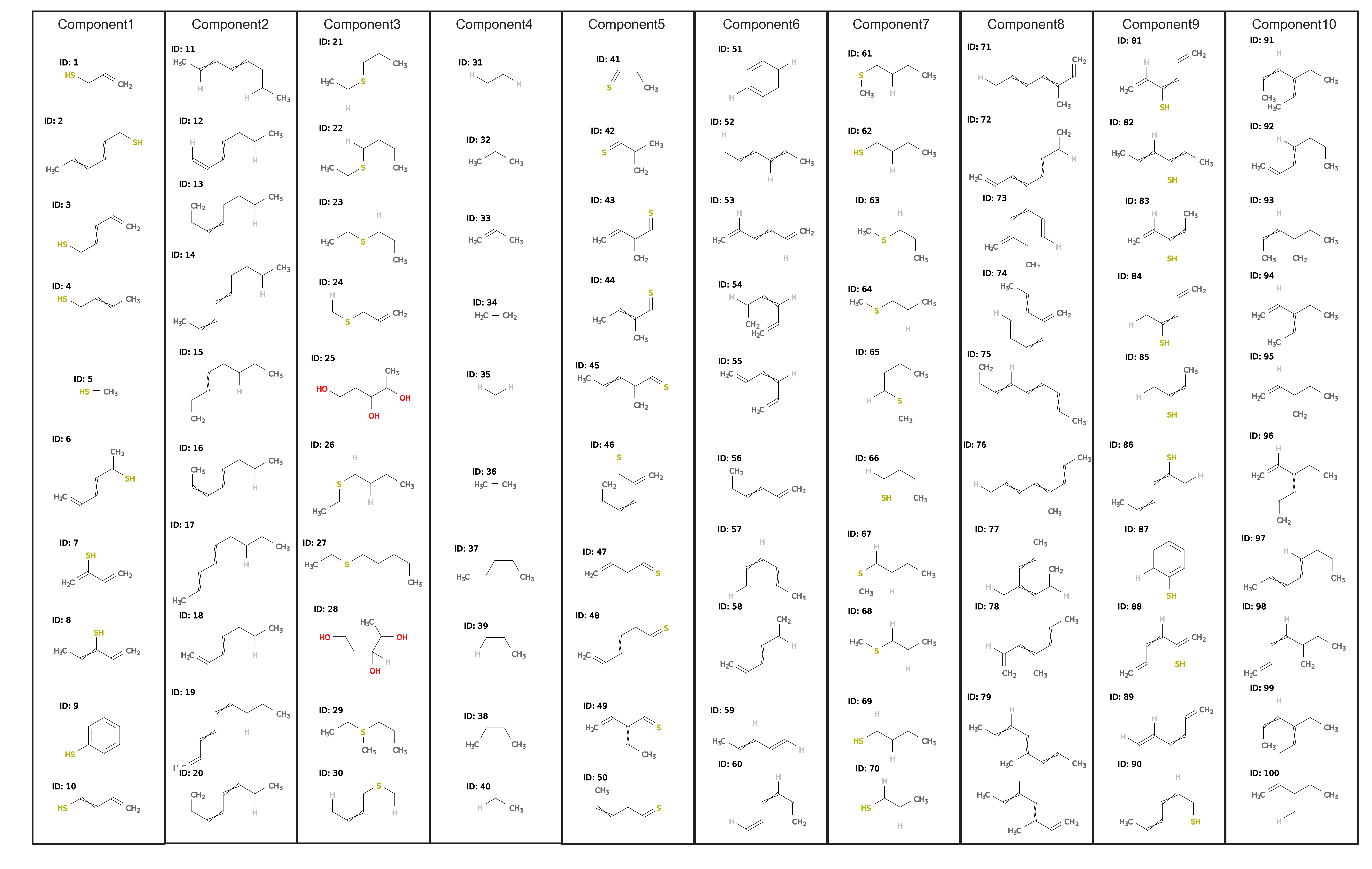}
\end{tabular}
\end{center}
\vspace{-0.8cm}
\caption{Extracted features for the top 10 latent components in the application of cPLS to the Compound dataset. Each column represents the highly weighted features (chemical substructures) of a latent component.}
\label{fig:com}
\end{figure*}

\begin{table*}
\caption{Compression size in mega bytes (MB), compression time in seconds (sec), and working space for hash table (MB) for varying parameter $\ell$ in Lossy-Re-Pair and $v$ in Freq-Re-Pair for each dataset.}
\begin{center}
{\footnotesize
  \begin{tabular}{r|cccc|cccc}
\hline
\multicolumn{1}{c}{} &\multicolumn{8}{c}{{\bf Book-review}}  \\
\hline
\multicolumn{1}{c}{} &\multicolumn{4}{|c|}{{\bf Lossy-RePair} $\ell$(MB)} & \multicolumn{4}{c}{{\bf Freq-RePair} $v$(MB)}  \\
\hline
 & 1 & 10 & 100 & 1000 & 1 & 10 & 100 & 1000\\
\hline
compression size~(MB) & 1836 & 1685 & 1502 & 1501 & 2021 & 1816 & 1680 & 1501 \\
compression time~(sec) & 9904 & 12654 & 19125 & 20004 & 2956 &2355 & 3165 & 21256 \\
working space (MB) & 1113 & 1931 & 7988 & 8603 & 292 & 616 & 3856 & 6724 \\
\hline\hline
\multicolumn{1}{c}{} & \multicolumn{8}{c}{{\bf Compound}} \\
\hline
\multicolumn{1}{c}{} &\multicolumn{4}{|c|}{{\bf Lossy-RePair} $\ell$(MB)} & \multicolumn{4}{c}{{\bf Freq-RePair} $v$(MB)} \\
\hline
 & 1 & 10 & 100 & 1000 & 1 & 10 & 100 & 1000\\
\hline
compression size~(MB) & 1288 & 859 & 825 & 825 & 1523 & 1302 & 825 & 825 \\
compression time~(sec) & 5096 & 7053 & 7787 & 7946 & 1362 & 1587 & 8111 & 8207 \\
working space~(MB) & 1113 & 1926 & 5030 & 5030 & 292 & 616 & 3535 & 3535 \\
\hline\hline

\multicolumn{1}{c}{} & \multicolumn{8}{c}{{\bf Webspam}} \\
\hline
\multicolumn{1}{c}{} &\multicolumn{4}{|c|}{{\bf Lossy-RePair} $\ell$(MB)} & \multicolumn{4}{c}{{\bf Freq-RePair} $v$(MB)}  \\
\hline
 & 1 & 10 & 100 & 1000 & 1 & 10 & 100 & 1000\\
\hline
compression size~(MB) & 1427 & 948 & 940 & 940 & 2328 & 2089 & 1050 & 940 \\
compression time~(sec) & 6953 & 10585 & 10584 & 10964 & 2125 & 2799 & 7712 & 11519  \\
working space~(MB) & 1112 & 1923 & 7075 & 7075 & 292 & 616 & 3856 & 5539 \\
\hline\hline
\multicolumn{1}{c}{} &\multicolumn{8}{c}{{\bf CP-interaction}} \\
\hline
\multicolumn{1}{c}{} &\multicolumn{4}{|c|}{{\bf Lossy-RePair} $\ell$(MB)} & \multicolumn{4}{c}{{\bf Freq-RePair} $v$(MB)}  \\
\hline
 & 10 & 100 & 1000 & 10000 & 10 & 100 & 1000 & 10000 \\
\hline
compression size~(MB) & - & 5199  & 5139 & 5036 & 20307 & 9529 & 5136 & 5136 \\
compression time~(sec) & 24hours & 55919 & 44853 & 43756 & 24565 & 39647 & 47230 & 48653 \\
working space~(MB)    & - & 9914  & 16650 & 16635 & 616 & 3856 & 13796 & 13796 \\
\hline\hline
\multicolumn{1}{c}{} & \multicolumn{8}{c}{{\bf CP-intensity}} \\
\hline
\multicolumn{1}{c}{} &\multicolumn{4}{|c|}{{\bf Lossy-RePair} $\ell$(MB)} & \multicolumn{4}{c}{{\bf Freq-RePair} $v$(MB)}  \\
\hline
      & 10   & 100 & 1000  & 10000 & 10 & 100 & 1000 & 10000  \\
\hline
compression size~(MB) & 558  & 543  & 540  & 535 & 588 & 535 & 535 & 535 \\
compression time~(sec) & 8103 & 6479 & 6494 & 6657 & 5423 & 5848 & 5859 & 5923 \\
working space~(MB)    & 1936 & 3552 & 3722 & 3738 & 616 & 2477 & 2477 & 2477 \\
\hline
  \end{tabular}
}
\end{center}
\label{tab:res2}

\caption{Results of cPLS, PCA-SL, bMH-SL and SGD for various datasets.  
         Dspace: the working space for storing data matrix (MB), Ospace: the working space for optimization algorithm and Ltime: learning time (sec). }
\vspace{-0.2cm}
{\footnotesize
\begin{center}
  \begin{tabular}{r|c|c|c|c|c|c}
\multicolumn{1}{c|}{} &  \multicolumn{5}{c|}{{\bf cPLS}} \\
\hline
Data               & $m$  & Dspace(MB) & Ospace(MB) & Ltime(sec) & AUC/PCC \\
\hline
Book-review    & 100  & 1288     & 15082 & 21628  &  0.96 \\
Compound       & 20   & 786      & 7955  & 1089   &  0.83 \\
Webspam        & 60   & 890      & 7736  & 4171   &  0.99 \\
CP-interaction & 40   & 4367     & 53885 & 35880  &  0.77 \\
CP-intensity   & 60   & 472      & 10683 & 33969  &  0.67 \\
\hline
\multicolumn{1}{c|}{} &  \multicolumn{5}{c|}{{\bf PCA-SL}} \\
\hline
Data              & $m/C$  &  Dspace(MB)  & Ospace(MB) & Ltime(sec) & AUC/PCC \\
\hline
Book-review    & 100/1   &  14747 & 110    & 6820     & 0.70 \\
Compound       & 25/0.1 &  12     &  1     & 6        & 0.65 \\
Webspam        & 50/1   &  200    &  2     & 129      & 0.99 \\
CP-interaction &  -     &  -      &  -     & >24hours    &  -  \\
CP-intensity   & 100/0.1   &  1521    &  11    & 42       & 0.11 \\
\hline
\multicolumn{1}{c|}{} &  \multicolumn{5}{c|}{{\bf bMH-SL}} \\
\hline
Data               & $b/h/C$     & Dspace(MB) & Ospace(MB) & Ltime(sec) & AUC/PCC \\
\hline
Book-review    & 100/16/0.01 & 2457  & 110    & 1033  &  0.95 \\
Compound       & 30/16/10    & 2     & 1      & 1     &  0.62 \\
Webspam        & 30/16/10    & 20    & 2      & 2     &  0.99 \\
CP-interaction & 30/16/0.1   & 12366 & 1854   & 10054 &  0.77 \\
CP-intensity   & 100/16/0.1  & 253   & 11     & 45    &  0.54 \\
\hline
\multicolumn{1}{c|}{} &  \multicolumn{5}{c|}{{\bf SGD}} \\
\hline
Data              & $C$  &  Dspace(MB)  & Ospace(MB) & Ltime(sec) & AUC/PCC \\
\hline
Book-review    & 10   &    -      & 1694     &  57   &  0.96 \\
Compound       & 10   &    -      & 9539     &  83   &  0.82 \\
Webspam        & 10   &    -      & 3041     &  85   &  0.99 \\
CP-interaction & 1    &    -      & 663      & 3163  &  0.75 \\
CP-intensity   & 0.1  &    -      & 124      & 280   &  0.04 \\
\hline
  \end{tabular}
\end{center}
}
\label{tab:acc}
\end{table*}

In this section, we demonstrate the effectiveness of cPLS with massive datasets. 
We used five datasets, as shown in Table~\ref{tab:statistics}.
"Book-review" consists of 12,886,488 book reviews in English from Amazon~\cite{McAuley13}. 
We eliminated stop-words from the reviews and then represented  
them as 9,253,464 dimensional fingerprints, where each dimension of the fingerprint represents the presence or absence of a word. 
"Compound" is a dataset of 42,682 chemical compounds that are represented as labeled graphs. 
We enumerated all the subgraphs of at most 10 vertices from the chemical graphs by using gSpan~\cite{Yan02} and 
then converted each chemical graph into a 52,099 dimensional fingerprint, where each dimension of the fingerprint represents the presence or absence of a chemical substructure. 
"Webspam" is a dataset of 16,609,143 fingerprints of 350,000 dimensions\footnote{The dataset is downloadable from \url{http://www.csie.ntu.edu.tw/~cjlin/libsvmtools/datasets/binary.html}.}.
"CP-interaction" is a dataset of 216,121,626 compound-protein pairs, where each compound-protein pair is represented as a 3,621,623 dimensional fingerprint and 300,202 compound-protein pairs are interacting pairs according to the STITCH database~\cite{Kuhn10}. 
We used the above four datasets for testing the binary classification ability. 
"CP-intensity" consists of 1,329,100 compound-protein pairs represented as 682,475 dimensional fingerprints, where the information about compound-protein interaction intensity was obtained from several chemical databases (e.g., ChEMBL, BindingDB and PDSP Ki). 
The intensity was observed by IC50 (half maximal (50\%) inhibitory concentration).
We used the "CP-intensity" dataset for testing the regression ability.
The number of all the nonzero dimensions in each dataset is summarized in the \#nonzero column in Table~\ref{tab:statistics}, and the size for storing fingerprints in memory by using 32bits for each element is written in the memory column in Table~\ref{tab:statistics}. 
We implemented all the methods by C++ and performed all the experiments on one core of a quad-core Intel Xeon CPU E5-2680 (2.8GHz). 
We stopped the execution of each method if it had not finished within 24hours in the experiments. 
In the experiments, cPLS did not use a secondary storage device for compression, i.e., cPLS compressed data matrices by loading all data in memory. 

\subsection{Compression Ability and Scalability}\label{sec:comp}
First, we investigated the influence on compression performance of the top-$k$ parameter in our Re-Pair algorithms.
For this setting, we used the Lossy-Re-Pair algorithm, where parameter $\ell$ is set to the total length of all rows in an input data matrix in order to keep all the symbols in the hash table. 
We examined $k=\{1\times 10^4, 2.5\times 10^4, 5\times 10^4, 7.5\times 10^4, 10\times 10^4 \}$ for the Book-review, Compound and Webspam datasets and examined
$k=\{1\times 10^5, 2.5\times 10^5, 5\times 10^5, 7.5\times 10^5, 10\times 10^5\}$ for the CP-interaction and CP-intensity datasets. 

Figure~\ref{fig:topk} shows compression size and compression time for various top-$k$.
We observed a trade-off between compressed size and compression time for all the datasets. 
The smaller the compressed size, the larger the compression time for larger values of $k$. 
In particular, significantly faster compression time was possible at the cost of only slightly worse compression.
For example, Lossy-Re-Pair took 57,290 seconds to compress the Book-review dataset and its size was 1,498 mega bytes (MB) for $k$=10000. 
When $k$=100000, compression time dropped to 20,004 seconds (less than half), while compressed size increased negligibly to 1,502MB. 

The same trends for the Book-review dataset were observed in the other datasets, which suggests that in practice a large value of $k$ can be chosen for fast compression,
without adversely affecting compression performance. 
Notably, we observed our compression method to be particularly effective for the larger datasets: CP-interaction and CP-intensity. 
The original sizes of CP-interaction and CP-intensity were 125GB and 110GB, respectively, while the compressed sizes of CP-interaction and CP-intensity were at most 5GB and at 535MB, respectively.
Our compression method thus achieved compression rates of 4\% and less than 1\% for CP-interaction and CP-intensity, respectively. 
Such significant reductions in data size enable the PLS algorithm to scale to massive data.

Next, we evaluated the performance of Lossy-Re-Pair and Freq-Re-Pair, where parameters $\ell$$=$$\{1$MB, $10$MB, $100$MB, $1000$MB$\}$ were examined for Lossy-Re-Pair, and parameters $v$ $=$$\{1$MB, $10$MB, $100$MB, $1000$MB$\}$ and $\epsilon=\{30\}$ were examined for Freq-Re-Pair. 
Table~\ref{tab:res2} shows the compressed size, compression time and the working space used for the hash table in Lossy-Re-Pair and Freq-Re-Pair. 
We observed that both Lossy-Re-Pair and Freq-Re-Pair achieved high compression rates using small working space. 
Such efficiency is crucial when the goal is to compress huge data matrices that exceed the size of RAM; our Re-Pair algorithm can compress data matrices stored in external memory (disk).
For compressing the CP-interaction dataset, Lossy-Re-Pair and Freq-Re-Pair consumed 16GB and 13GB, respectively, achieving a compressed size of 5GB. 
We observed the same tendency for the other datasets (See Table~\ref{tab:res2}).

\subsection{Prediction Accuracy}\label{sec:pred}
We evaluated the classification and regression capabilities of cPLS, PCA-SL, bMH-SL and SGD. 
Following the previous works~\cite{Yu10, Li11}, we randomly selected 20\% of samples for testing and used the remaining 80\% of samples for training. 
cPLS has one parameter $m$, so we selected the best parameter value among $m=\{10, 20,...,100\}$ that achieved the highest accuracy for each dataset. 
The PCA phase of PCA-SL has one parameter deciding the number of principal components $m$, which was chosen from $m=\{10,25,50,75,100\}$ whose 
maximum value of $100$ is the same as that of cPLS's parameter $m$. 
Linear models were learned with LIBLINEAR~\cite{Fan08}, one of the most efficient implementations of linear classifiers, on PCA's compact feature vectors, where the hinge loss of linear SVM for classification and the squared error loss for regression were used with $L_2$-regularization.
The learning process of PCA-SL~\cite{Jolliffe86,Elgamal15} has one parameter $C$ for $L_2$-regularization, which was chosen from $C=\{10^{-5}, 10^{-4},$ $...,10^{5}\}$. 
For PCA-SL~\cite{Jolliffe86,Elgamal15}, we examined all possible combinations of two parameters ($m$ and $C$) and selected the best combination achieving the highest accuracy for each dataset. 
The hashing process of bMH-SL~\cite{Li11} has two parameters (the number of hashing values $h$ and the length of bits $b$), so we examined all possible combinations of $h=\{10, 30, 100\}$ and $b=\{8, 16\}$.
As in PCA-SL, linear models were learned with LIBLINEAR~\cite{Fan08} on bMH's compact feature vectors, where the hinge loss of linear SVM for classification and the squared error loss for regression were used with $L_2$-regularization.
The learning process of bMH-SL~\cite{Li11} has one parameter $C$ for $L_2$-regularization, which was chosen from $C=\{10^{-5}, 10^{-4},$ $...,10^{5}\}$. 
For bMH-SL, we examined all possible combinations of three parameters ($h$, $b$, and $C$) and selected the best combination achieving the highest accuracy for each dataset. 
We implemented SGD on the basis of the AdaGrad algorithm~\cite{Duchi11} using the logistic loss for classification and the squared error loss for regression with $L_2$-regularization.
SGD~\cite{Duchi11} has one parameter $C$ for $L_2$-regularization, which was also chosen from $C=\{10^{-5}, 10^{-4},...,10^{5}\}$. 
We measured the prediction accuracy by the area under the ROC curve (AUC) for classification and Pearson correlation coefficient (PCC) for regression. 
Note that AUC and PCC return 1 for perfect inference in classification/regression, while AUC returns 0.5 for random inference and PCC returns 0 for random inference.
We report the best test accuracy under the above experimental settings for each method below. 

Table~\ref{tab:acc} shows the prediction accuracy, working space and training time of cPLS, PCA-SL, bMH-SL, and SGD. The working space for the storing data matrix and the working space needed for optimizations were separately evaluated.
While PCA-SL and bMH-SL significantly reduced the working space for storing data matrices, 
the classification and regression accuracies were low. 
Since PCA-SL and bMH-SL compress data matrices without looking at the correlation structure 
between data and output variables for compressing data matrices, 
high classification and regression accuracies were difficult to achieve.


SGD significantly reduced the working space, since it did not store data matrices in memory. 
Classification and regression accuracies of SGD were not high, because of the instability of the optimization algorithm.
In addition, SGD is applicable only to simple linear models,
making the learned model difficult to interpret.

Our proposed cPLS outperformed the other methods (PCA-SL, bMH-SL, and SGD) in terms of AUC and PCC and significantly reduced the working space. 
The results showed cPLS's efficiency for learning PLS on compressed data matrices while looking at the correlation structure between data and output variables. 
Such a useful property enables us to extract informative features from the learned model.



\subsection{Interpretability}\label{seq:int}
Figure~\ref{fig:com} shows the top-$10$ highly-weighted features that were extracted for each component in the application of cPLS to the Compound dataset, where one feature corresponds to a compound chemical substructure.
It was observed that structurally similar chemical substructures were extracted together as important features in the same component, and
the extracted chemical substructures differed between components.
This observation corresponds to a unique property of cPLS.
Analysing large-scale compound structure data is of importance in pharmaceutical applications, especially for rational drug design.
For example, the extracted chemical substructures are beneficial for users who want to identify important chemical fragments involved in therapeutic drug activities or adverse drug reactions.

\section{Conclusions and Future Work}
We presented a scalable algorithm for learning interpretable linear models --- called cPLS --- which is applicable to large-scale regression and classification tasks. 
Our method has the following appealing properties: 
 \vspace{-0.5\baselineskip}           
{
\setlength{\leftmargini}{10pt}         
\begin{enumerate}
\setlength{\parskip}{0.02cm} 
\setlength{\itemsep}{0.02cm} 
\setlength{\itemindent}{0pt}   
\setlength{\labelsep}{3pt}     
\item {\bf Scalability}: cPLS is applicable to large numbers of high-dimensional fingerprints (see Sections~\ref{sec:comp} and ~\ref{sec:pred}).
\item {\bf Prediction Accuracy}: The optimization of cPLS is numerically stable, which enables us to achieve high prediction accuracies (see Section~\ref{sec:pred}). 
\item{\bf Usability}: cPLS has only one hyperparameter to be tuned in cross-validation experiments (see Section~\ref{sec:cpls}).
\item {\bf Interpretability}: Unlike lossy compression-based methods, cPLS can extract informative features reflecting the correlation structure between data and class labels (or response variables), which makes the learned models easily interpretable (see Section~\ref{seq:int}).
\end{enumerate} 
}
\vspace{-0.5\baselineskip}           


In this study, we applied our proposed grammar compression algorithm to scaling up PLS, but 
in principle it can be used for scaling up other machine learning methods or data mining techniques.
An important direction for future work is therefore the development of scalable learning methods and data mining techniques based on grammar-compression techniques. 
Such extensions will open the door for machine learning and data mining methods to be applied in various large-scale data problems in research and industry. 




\section{Acknowledgments}
This work was supported by MEXT/JSPS Kakenhi (24700140, 25700004 and 25700029), the JST PRESTO program, the Program to Disseminate Tenure Tracking System, MEXT and Kyushu University Interdisciplinary Programs in Education and Projects in Research Development, and the Academy of Finland via grant 294143.

\bibliographystyle{plain}
\bibliography{biblio}

\end{document}